\documentclass[epj]{svjour}
\usepackage{graphics}
\begin{document}

\newcommand{\vk}{{\vec k}}
\newcommand{\vK}{{\vec K}}
\newcommand{\vb}{{\vec b}}
\newcommand{{\vp}}{{\vec p}}
\newcommand{{\vq}}{{\vec q}}
\newcommand{\vQ}{{\vec Q}}
\newcommand{\vx}{{\vec x}}
\newcommand{\beq}{\begin{equation}}
\newcommand{\eeq}{\end{equation}}
\newcommand{\half}{{\textstyle \frac{1}{2}}}
\newcommand{\gton}{\stackrel{>}{\sim}}
\newcommand{\lton}{\mathrel{\lower.9ex \hbox{$\stackrel{\displaystyle<}{\sim}$}}}
\newcommand{\ee}{\end{equation}}
\newcommand{\ben}{\begin{enumerate}}
\newcommand{\een}{\end{enumerate}}
\newcommand{\bit}{\begin{itemize}}
\newcommand{\eit}{\end{itemize}}
\newcommand{\bc}{\begin{center}}
\newcommand{\ec}{\end{center}}
\newcommand{\bea}{\begin{eqnarray}}
\newcommand{\eea}{\end{eqnarray}}

\newcommand{\beqar}{\begin{eqnarray}}
\newcommand{\eeqar}[1]{\label{#1} \end{eqnarray}}
\newcommand{\pleft}{\stackrel{\leftarrow}{\partial}}
\newcommand{\pright}{\stackrel{\rightarrow}{\partial}}

\newcommand{\eq}[1]{Eq.~(\ref{#1})}
\newcommand{\fig}[1]{Fig.~\ref{#1}}
\newcommand{\eff}{ef\!f}
\newcommand{\alphas}{\alpha_s}

\renewcommand{\topfraction}{0.85}
\renewcommand{\textfraction}{0.1}
\renewcommand{\floatpagefraction}{0.75}

\title{Nuclear suppression of $\phi$ meson yields with large $p_T$ at the RHIC and the LHC}
\author{Wei Dai\inst{1} \and Xiao-Fang Chen\inst{2} \and Ben-Wei Zhang\inst{1} \thanks{ bwzhang@mail.ccnu.edu.cn} \and Han-Zhong Zhang\inst{1} \and Enke Wang\inst{1}
%
%
}                     
%
%
\institute{Key Laboratory of Quark \& Lepton Physics (MOE) and Institute of Particle Physics,
 Central China Normal University, Wuhan 430079, China \and School of Physics and Electronic Engineering, Jiangsu Normal University, Xuzhou 221116, China}

\date{Received: date / Revised version: date}
%
\abstract{
We calculate $\phi$ meson transverse momentum spectra in p+p collisions as well as their nuclear suppressions in central A+A collisions both at the RHIC and the LHC in LO and NLO with the QCD-improved parton model.   We have included the parton energy loss effect in hot/dense QCD medium with the effectively medium-modified $\phi$ fragmentation functions in the higher-twist approach of jet quenching. The
nuclear modification factors of $\phi$ meson in central Au+Au collisions at the RHIC and central Pb+Pb collisions at the LHC are provided, and a nice agreement of our numerical results at NLO with the ALICE measurement is observed. Predictions of yield ratios of neutral mesons such as $\phi/\pi^0$, $\phi/\eta$ and $\phi/\rho^0$ at large $p_T$ in relativistic heavy-ion collisions are also presented for the first time.
} 
\maketitle
\section{Introduction}
\label{intro}
Jet quenching phenomena~\cite{Wang:1991xy}, as one of the key discoveries made so far in relativistic heavy-ion collisions (HIC) at the RHIC and the LHC, have been extensively studied for a wide range of observables. Among them, the yield suppression of the produced final state hadrons at large transverse momentum $p_T$~\cite{Gyulassy:2003mc} provides the most direct and one of the fundamental observables which can reveal the mechanism of parton energy loss in dense QCD medium and test  many-body QCD theory. Experimentally, a suppression of approximately same magnitude is observed for $\pi^0$, $\eta$ and $\rho^0$ productions at the RHIC despite of their different masses~\cite{Adler:2006hu,Adare:2010cy,Agakishiev:2011dc}, and a similar observation has also be made by ALICE Collaboration~\cite{Morreale:2016dli}.
By considering the fast parton suffers medium induced energy loss while propagating through the QCD medium before its fragmenting into final state hadrons in the vacuum outside the QCD medium,
we explore the suppression pattern of these neutral mesons with the next-to-leading order (NLO) calculations in the QCD-improved parton model  in the previous publications~\cite{Chen:2010te,Chen:2011vt,Dai:2015dxa,Dai:2017tuy}. It has been demonstrated that for productions of $\pi^0$, $\eta$ and $\rho^0$ mesons containing light valence quark,  quark fragmentation gives the largest contributions at large $p_T$; with the relatively weak $p_T$ and $z_h$ (momentum fraction of partons carried by the fragmentated hadrons) dependence of their quark fragmentation functions (FFs), even though jet quenching effect will alter the gluon and quark relative contributions to the yields of these neutral mesons in HIC relative to in p+p collisions, the $\eta/\pi^0$ and $\rho^0/\pi^0$ ratios in HIC will eventually coincide with that in p+p at very large $p_T$~\cite{Dai:2015dxa,Dai:2017tuy}. 

In this paper, we apply the same framework to investigate $\phi(s{\bar s})$ meson, which is also a light meson but contains strange (anti-stange) valence quarks. The production of $\phi$ meson can be used to probe different aspects of the heavy ion collisions, such as the strangeness enhancement and chiral symmetry restoration. Here, we focus on parton energy loss effect on $\phi$ meson cross section at large $p_T$ and its nuclear modification factor to further examine the particle species dependence of jet quenching in the QCD medium. We notice that due to the lack of precise parametrizations of fragmentation functions of $\phi$ meson, theoretical calculations of the $\phi$ meson production at large $p_T$ in either p+p collisions or HIC  at the RHIC and the LHC have not been available so far.

To make a perturbative QCD calculation of $\phi$ production at large transverse momentum, the parton FFs of $\phi$ meson   $D_{q,g\rightarrow \phi}(z_h,Q)$ at any hard scale $Q$ should be needed.
In our study
we utilize  the availability of an broken SU(3) model description of vector mesons productions~\cite{Saveetha:2013jda,Indumathi:2011vn} to have an initial parametrization of $\phi$ FFs in vacuum at a starting energy scale $\rm Q_{0}^2=1.5\ GeV^2$ as an input. 
We calculate the productions of $\phi$ meson in p+p collision at $\sqrt{s_{NN}}=200$~GeV 
and $\sqrt{s_{NN}}=2.76$~TeV up to NLO, and find the theoretical results to be in good agreement with the PHENIX and ALICE data respectively. Then we numerically investigate the $\phi$ meson productions in A+A collisions by incorporating the effectively medium modified FFs in the higher twist approach of parton energy loss. We provide
for the first time the numerical results of the $\phi$ meson yields in central A+A collisions both at the RHIC and the LHC. We confront the theoretical results of the nuclear modification factor $R_{\rm AA}(\phi)$ in Pb+Pb collisions at LHC with the existing experimental data by ALICE Collaboration, and find they match well with each other.  We may explore further how the change of jet chemistry due to 
partonic energy loss results in  different suppression patterns between $\phi(s\bar{s})$ meson and other neutral mesons such as $\pi^0$, $\eta$ and $\rho^0$ by plotting yield ratios of  $\phi/\pi^0$, $\phi/\eta$ and  $\phi/\rho^0$ in p+p and in HIC.

\section{Large $p_T$ yield of $\phi$ meson in p+p}
\label{p+p}
The leading hadron production in p+p collisions can be factorized into three parts as parton distribution functions (PDFs) inside the incoming protons, elementary partonic scattering cross sections $d\hat{\sigma}/d\hat{t}$, and parton FFs to the final state hadron~\cite{Owens:1986mp}. 
To facilitate the discussions of parton FFs in vacuum and medium, we take the following formula:
\begin{eqnarray}
\frac{1}{p_{T}}\frac{d\sigma_{h}}{dp_{T}}=\int F_{q}(\frac{p_{T}}{z_{h}})\cdot D_{q\to h}(z_{h}, Q^2)\frac{dz_{h}}{z_{h}^2} \nonumber  \\
+ \int F_{g}(\frac{p_{T}}{z_{h}})\cdot D_{g\to h}(z_{h}, Q^2)\frac{dz_{h}}{z_{h}^2}  \,\,\, .
\label{eq:ptspec}
\end{eqnarray} 
One can see the single hadron production in $\rm p+p$ collision will be determined by two factors: the initial hard (parton-)jet spectrum $F_{q,g}(p_T)$ and the parton FFs $D_{q,g\to h}(z_{h}, Q^2)$ to the final-state hadron.  In our calculations, we employed CTEQ6M parametrization for PDFs~\cite{Lai:1999wy} in colliding protons, which has been convoluted with partonic scattering cross sections $d\hat{\sigma}/d\hat{t}$ to obtain $F_{q,g}(\frac{p_{T}}{z_{h}})$. $D_{q,g\to h}(z_{h}, Q^2)$ denotes the parton FFs in vacuum, which give the possibilities of scattered parton fragmenting into hadron $h$ at momentum fraction $z_h$ and fragmentation scale $Q$. In practice,
the factorization, renormalization and fragmentation scales are usually chosen to be the same and proportional to the final-state $\rm p_T$ of the leading hadron.

Due to the paucity of the experimental data, there are few parameterized $\phi$ parton FFs.  Recently a broken $SU(3)$ model is proposed to extracting  parton FFs of the vector mesons~\cite{Saveetha:2013jda,Indumathi:2011vn}. The complexity of the meson octet fragmentation functions has been reduced considerably by introducing the $SU(3)$ flavor symmetry with a symmetry breaking parameter. The isospin and charge conjugation invariance of the vector mesons further reduce independent quark flavor FFs into functions named valence(V) and sea($\gamma$). The inputs of valence $V(x, Q_0^2)$, sea $\gamma(x, Q_0^2)$ and gluon $D_g(x,Q_0^2)$ FFs are parameterized into a standard polynomial at a starting low energy scale of $Q_0^2=1.5$~$\rm GeV^2$ such as:
\begin{eqnarray}
F_i(x)=a_ix^{b_i}(1-x)^{c_i}(1+d_ix+e_ix^2)
\end{eqnarray}
In addition, since $\phi$ meson is dominated by its $s\bar{s}$ component, the FFs can be expressed as orthogonal combinations of the $SU(3)$ octet ($\omega_8$) and singlet states($\omega_1$):
\begin{eqnarray}
\phi=\cos\theta  \omega_8 - \sin \theta  \omega_1 \ .
\end{eqnarray}
And a few additional parameters such as $f_1^{u}$, $f_1^{s}$, $f_{sea}$ representing the singlet constants and the sea suppression, the vector mixing angle $\theta$ mentioned in the above equation are introduced. Together with the three sets of parameters in $V(x, Q_0^2)$, $\gamma(x, Q_0^2)$ and $D_g(x,Q_0^2)$, all these parameters are initially parameterized at starting scale of $Q^2=1.5$~$\rm GeV^2$ by evolving  through DGLAP equation~\cite{Hirai:2011si}, then fitting the cross section at NLO with the measurements of LEP($\rho$,$\omega$) and SLD($\phi$,$K^\star$) at $\sqrt{s}=91.2$~GeV. The parameters for $\phi$ FFs in vacuum at $Q^2=1.5$~$\rm GeV^2$ are listed in Ref.~\cite{Saveetha:2013jda,Indumathi:2011vn} and we obtain the $\phi$ meson FFs at any energy scale $Q$ by  evolving them in the numerical DGLAP equation at NLO~\cite{Hirai:2011si}.

\begin{figure}[!t]
\begin{center}
\hspace*{-0.1in}
\vspace*{-0.1in}
\resizebox{0.5\textwidth}{!}{%
\includegraphics{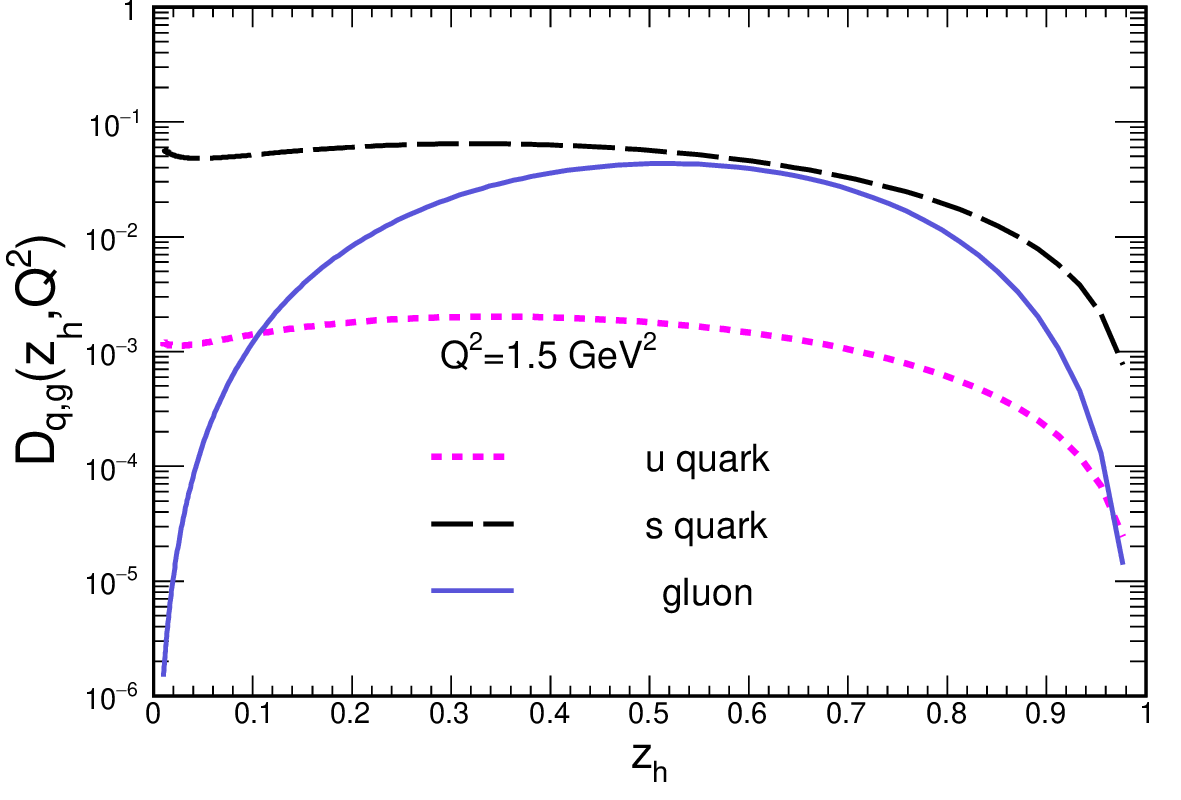}
\includegraphics{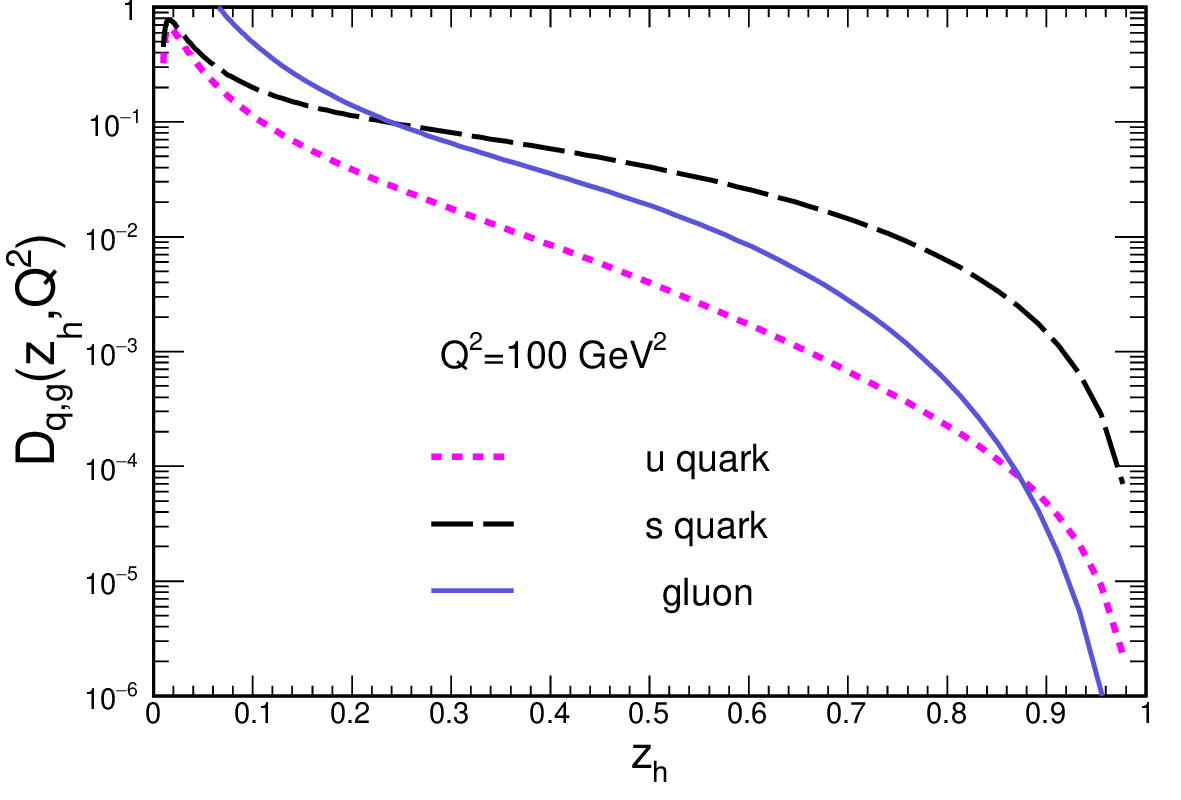}
}

\hspace*{-0.1in}
\vspace*{0.0in}
\caption{ Parton FFs of $\phi$ meson as a function of $z_h$ at the initial scale $Q^2=1.5\ \rm GeV^2$ (left panel), and at the scale of $Q^2=100\ \rm GeV^2$ (right panel).}
\label{fig:ffscale}
\end{center}
\end{figure}


\begin{figure}[!t]
\begin{center}
\hspace*{-0.1in}
\vspace*{-0.1in}
\resizebox{0.5\textwidth}{!}{%
\includegraphics{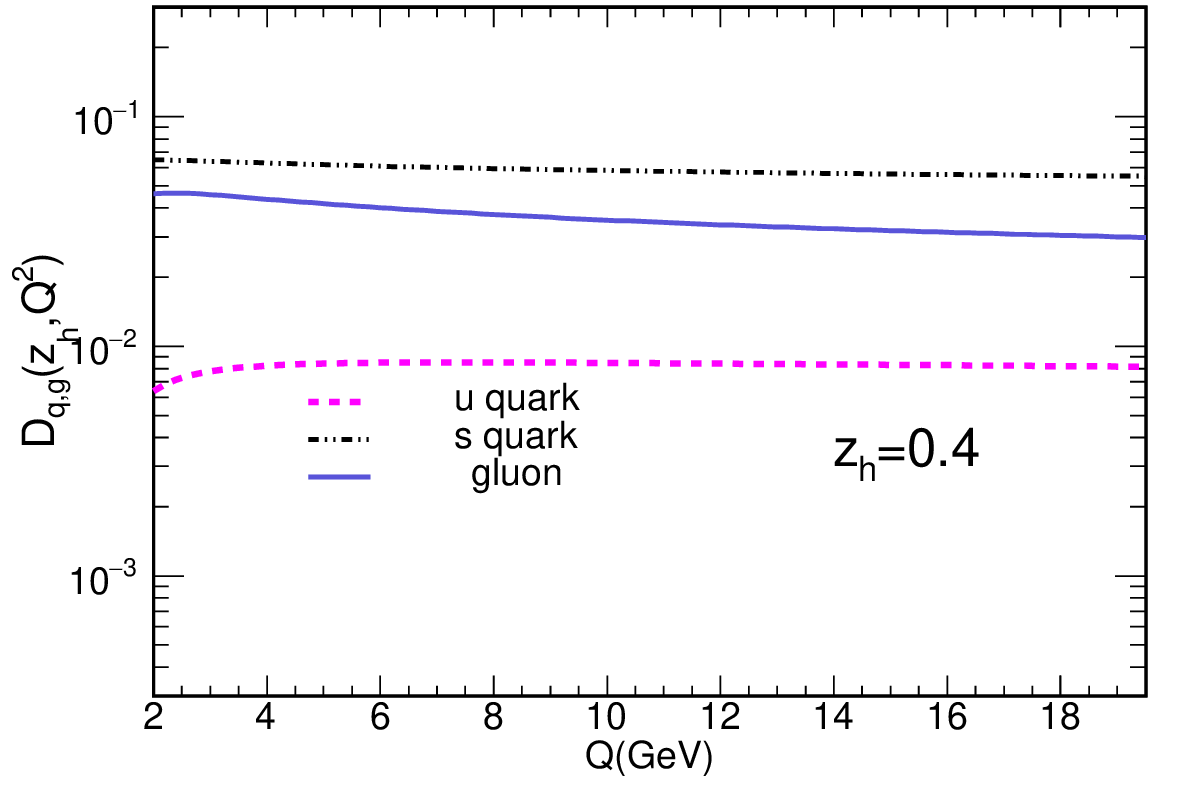}
\includegraphics{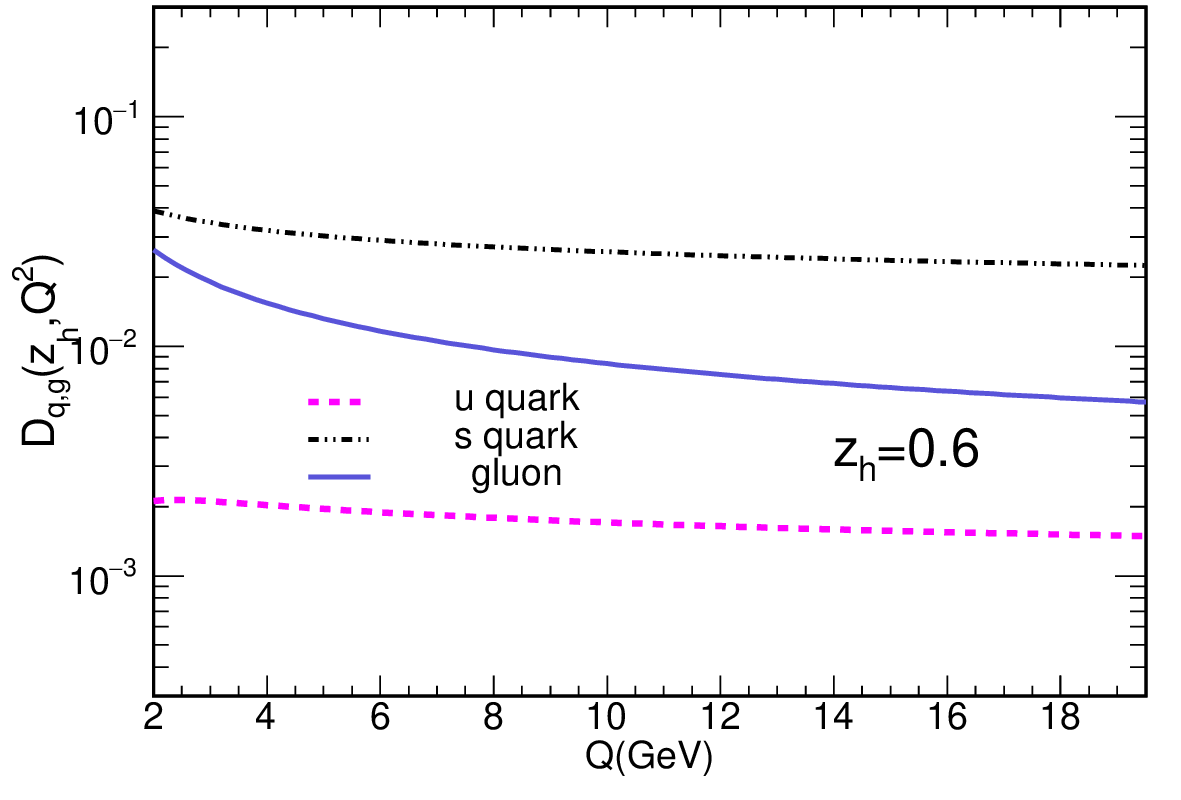}
}
\hspace*{-0.1in}
\vspace*{0.0in}
\caption{ Left: $\phi$ FFs as a function of the scale $Q$ at $z_h=0.6$; 
Right:   $\phi$ FFs as a function of the scale $Q$ at $z_h=0.6$.}
\label{fig:ffzh}
\end{center}
\end{figure}

To understand the pure $s\bar{s}$ state nature of $\phi$ and its influence to the production, We plot in Fig.~\ref{fig:ffscale} the initial parameterized FFs at initial energy scale $\rm Q^2=1.5\ GeV^2$ (left panel) and the DGLAP evolved FFs at $\rm Q^2=100\ GeV^2$ (right panel). It is observed that $D_{s\to \phi}(z_{h}, Q)\gg D_{u\to \phi}(z_{h}, Q) $, and in the intermediate and large $z_h$ regions gluon FF to $\phi$ meson is much larger than up (down) quark FFs to $\phi$. We notice these features are quite different from  FFs of other neutral mesons (such as $\rho^0$~\cite{Dai:2017tuy})
where up (down) quark FFs are much larger than strange quark FFs. In Fig.~\ref{fig:ffzh} we show the $Q$ ($p_T$) dependence of $\phi$ meson FFs at different fixed $z_h$. We can see in the plotted $Q$ region $D_{s\to \phi}(z_{h}, Q)> D_{g\to \phi}(z_{h}, Q) \gg D_{u\to \phi}(z_{h}, Q) $. Because in the initial hard scattering processes more gluon partons will be produced than strange quarks we may expect that there will be a competition between strange quark and gluon fragmentated contributions of the $\phi$ meson yield in p+p collision.

\begin{figure}[!t]
\begin{center}

\resizebox{0.5\textwidth}{!}{%
\includegraphics{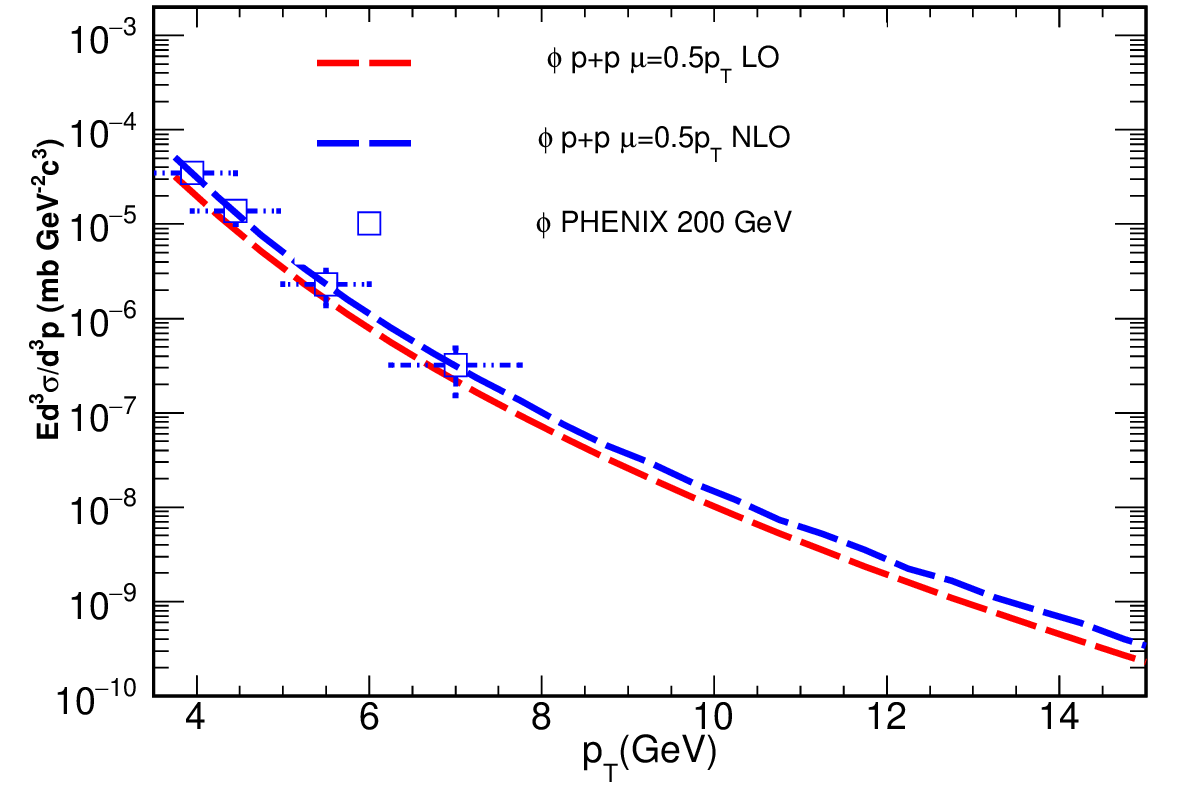}
}
\resizebox{0.5\textwidth}{!}{%
\includegraphics{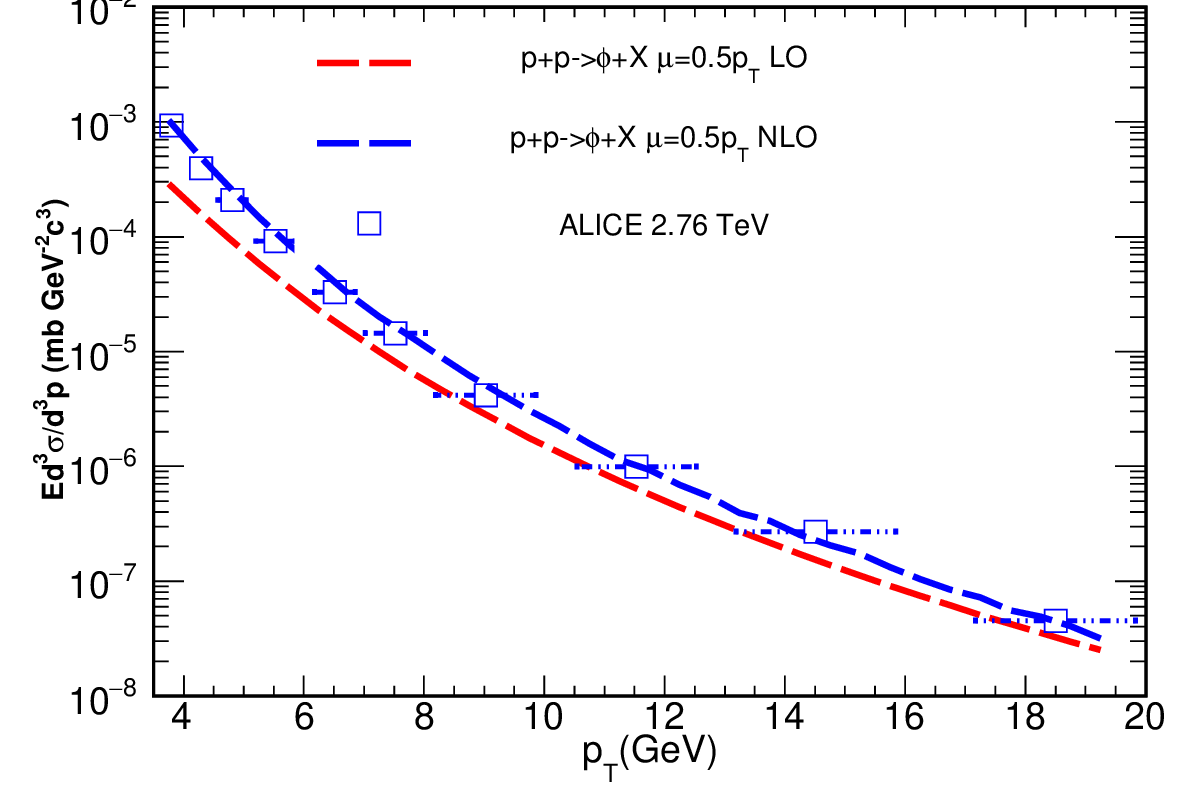}
}
\hspace*{-0.1in}
\caption{ Top: Numerical calculation of the $\phi$ production in $\rm p+p$ collisions with $\sqrt{s_{NN}}=200$~GeV at the RHIC with STAR data ~\cite{Adare:2010pt}($\sigma_{\rm NN}=42$~mb). Bottom: Numerical calculation of the $\phi$ production in $\rm p+p$ collisions with $\sqrt{s_{NN}}=2.76$~TeV at the LHC with ALICE data~\cite{Richer:2015vqa,Adam:2017zbf} ($\sigma_{\rm NN}=65$~mb~\cite{dEnterria:2003xac}). }
\label{fig:illustphipp}
\end{center}
\end{figure}

With the availability of $\phi$ FFs in vaccum, we make perturbative calculation of $\phi$ meson $p_T$ distribution in elementary proton-proton collisions at $\sqrt{s_{NN}}=200$~GeV and $\sqrt{s_{NN}}=2.76$~TeV with the QCD-improved parton model. In Fig.~\ref{fig:illustphipp} we 
 confront the numerical simulations at LO and NLO  with the experimental data of the $\phi$ production in p+p collision at the RHIC by STAR~\cite{Adare:2010pt} (top panel), and the LHC by ALICE~\cite{Adam:2017zbf} (bottom panel). One can see the NLO results with factorization and normalization scales $\rm \mu=0.5p_T$ give very nice description of data on $\phi$ cross section in p+p. In the following calculations the same hard scales  $\rm \mu=0.5p_T$ will be adopted. 

\section{Large $p_T$ yield of $\phi$ meson in HIC}
\label{HIC}

To study the single hadron productions in high-energy nuclear collisions, we have utilized the generalized factorization of twist-four processes to calculate parton energy loss due to medium-induced gluon radiation of a hard partons passing through the hot/dense QCD medium, and derive the effectively medium modified fragmentation functions in the higher-twist approach of parton energy loss~\cite{Guo:2000nz,Zhang:2003yn,Schafer:2007xh}. The effectively medium modified FF, which have effectively taken into account partonic energy loss effect, and used in the numerical simulations of leading hadron productions in A+A collisions, are written as~\cite{Chen:2010te,Chen:2011vt,Dai:2015dxa,Dai:2017tuy}:
\begin{eqnarray}
\tilde{D}_{q}^{h}(z_h,Q^2) &=&
D_{q}^{h}(z_h,Q^2)+\frac{\alpha_s(Q^2)}{2\pi}
\int_0^{Q^2}\frac{d\ell_T^2}{\ell_T^2} \nonumber\\
&&\hspace{-0.7in}\times \int_{z_h}^{1}\frac{dz}{z} \left[ \Delta\gamma_{q\rightarrow qg}(z,x,x_L,\ell_T^2)D_{q}^h(\frac{z_h}{z})\right.
\nonumber\\
&&\hspace{-0.2 in}+ \left. \Delta\gamma_{q\rightarrow
gq}(z,x,x_L,\ell_T^2)D_{g}^h(\frac{z_h}{z}) \right] .
\label{eq:mo-fragment}
\end{eqnarray}
which take a similar form to the vacuum bremsstrahlung corrections that leads to the DGLAP evolution for FFs in vacuum, with  the vacuum splitting functions replaced by the medium modified splitting functions $\Delta\gamma_{q\rightarrow qg}$ and $\Delta\gamma_{q\rightarrow gq}$. Therefore, to calculate the production of leading hadrons in A+A collision at the NLO,  we utilize the NLO partonic cross sections the same as in p+p, and the NLO nuclear PDFs, which are then convoluted with an effective medium-modified fragmentation function given by Eq.~(\ref{eq:mo-fragment}), where the vacuum FFs is evolved with NLO DGLAP equation while the correction convolutes a medium-induced kernel with the (DGLAP) evolved FFs at scale $Q^2$. The medium modified splitting functions depend on the twist-four quark-gluon correlations inside the medium $T^{A}_{qg}(x,x_L) $ which demonstrated by~\cite{Guo:2000nz,Zhang:2003yn}:
\begin{eqnarray}
\Delta\gamma_{q\rightarrow qg}(z,x,x_L,\ell_T^2)&=&
[\frac{1+z^2}{(1-z)_{+}}T_{qg}^{A}(x,x_L) \nonumber\\
&&\hspace{-0.7in} +\delta(1-z)\Delta
T_{qg}^{A}(x,x_L)]\frac{2\pi\alpha_sC_A}{\ell_T^2N_cf_q^A(x)};
\label{eq:delta-gamma} \\
\Delta\gamma_{q\rightarrow gq}(z,x,x_L,\ell_T^2)&=&\Delta\gamma_{q
\rightarrow qg}(1-z,x,x_L,\ell_T^2) \ .
 \label{eq:equal}
\end{eqnarray}
Due to the fact that the twist-four quark-gluon correlations $T^{A}_{qg}(x,x_L)$, which depend on the properties of the medium, can not be determined directly by the theoretical calculation.  By assuming a thermal ensemble of quasi-particle states in the hot and dense medium, and also neglect the multiple particle correlations inside the hot medium, we may have the quark-gluon correlation function in the higher-twist approach to multiple scattering in the QCD medium factorized as:~\cite{Chen:2010te,Chen:2011vt,Dai:2015dxa,Dai:2017tuy,Liu:2015vna}:
\begin{eqnarray}
\frac{T^{A}_{qg}(x,x_L)}{f_q^A(x)} &=&\frac{N_{c}^{2}-1}{4\pi\alpha_sC_{R}}\frac{1+z}{2} \int dy^{-}
2 \sin^{2}\left[\frac{y^{-}\ell_{T}^{2}}{4Ez(1-z)}\right]
\nonumber \\
&&\hspace{0.0 in}\times\left[\hat{q}_R(E,x_L,y)+c(x,x_{L}) \hat{q}_R(E,0,y)\right] \, 
 \label{eq:corr2}
\end{eqnarray}
Considering the contribution of the radiative energy loss and assuming $x \gg x_L, x_{T}$, we will have the jet transport parameter $\hat{q}_{R}(E,y) \equiv \hat{q}_{R}(E,x_L,y)\approx \hat{q}_{R}(E,0,y)$. Phenomenological given the evolutionary space and time profile to the jet transport parameter $\hat{q}_{R}(E,y)$, one can finally calculate the effective medium modified quark fragmentation function according to the Eq.~(\ref{eq:mo-fragment}).  The space-time evolution of the medium phenomenological is introduced by the value of jet transport parameter $\hat{q}$ relative to the initial value   $q_0$, located at the center of the overlap region at initial time of the QGP formation. We note the treatment here is model-dependent and a satisfactory treatment of medium-modifications of parton fragmentation from the first principle is still needed. To consider the radial flow, we also include the product of the four momentum of the jet and the four flow velocity of the medium along the jet propagation path in the collision frame~\cite{Chen:2011vt}.

The total energy loss embodied in the medium modified quark fragmentation function is the energies carried away by the radiative gluon (reflected by the medium modified splitting functions): 
\begin{eqnarray}
\frac{\Delta E}{E} &=& \frac{2N_{c}\alpha_s}{\pi} \int dy^-dz
{d\ell_T^2}
\frac{1+z^2}{\ell_T^4} \nonumber \\
&& \hspace{-0.5in}\times \left(1-\frac{1-z}{2}\right)\hat q(E,y)
\sin^2\left[\frac{y^-\ell_T^2}{4Ez(1-z)}\right], 
\label{eq:de-twist}
\end{eqnarray}
which is also  proportional to jet transport parameter $\hat{q}$. 

\hspace{0.7in}
\begin{figure}[!t]
\begin{center}

\resizebox{0.5\textwidth}{!}{%
\includegraphics{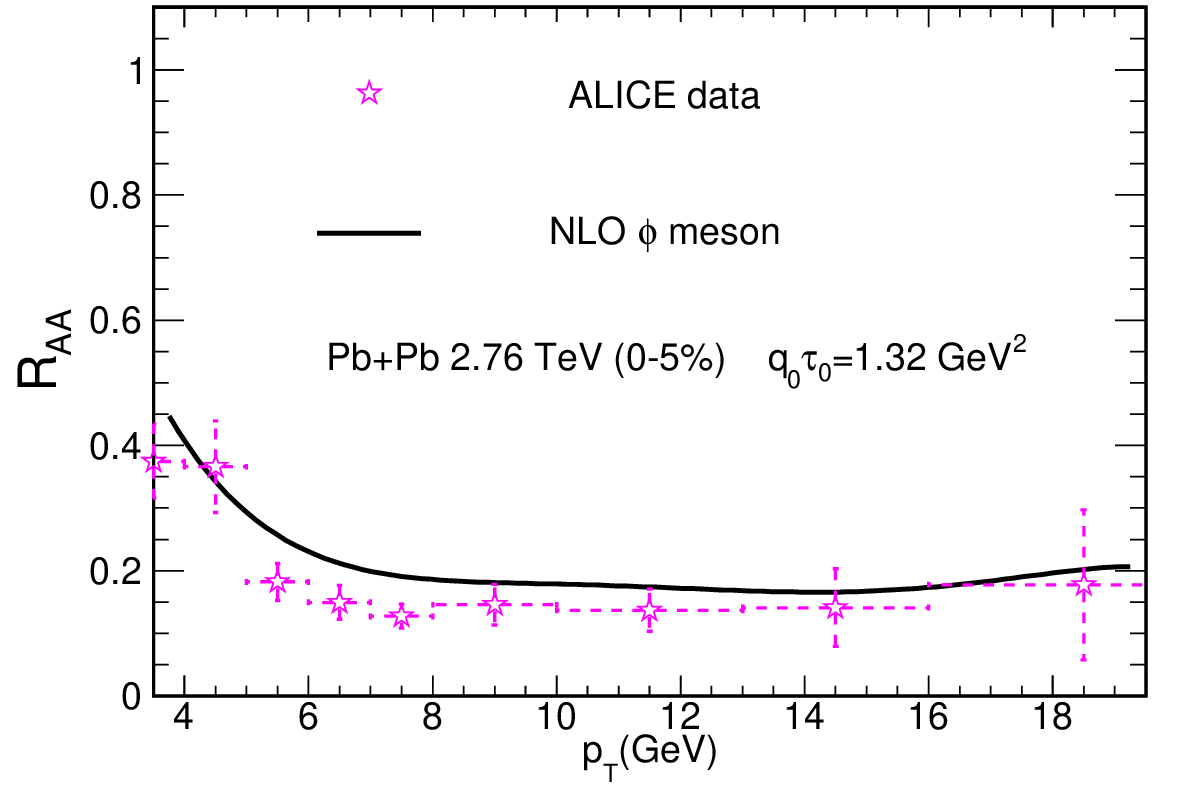}
}
\resizebox{0.5\textwidth}{!}{%
\includegraphics{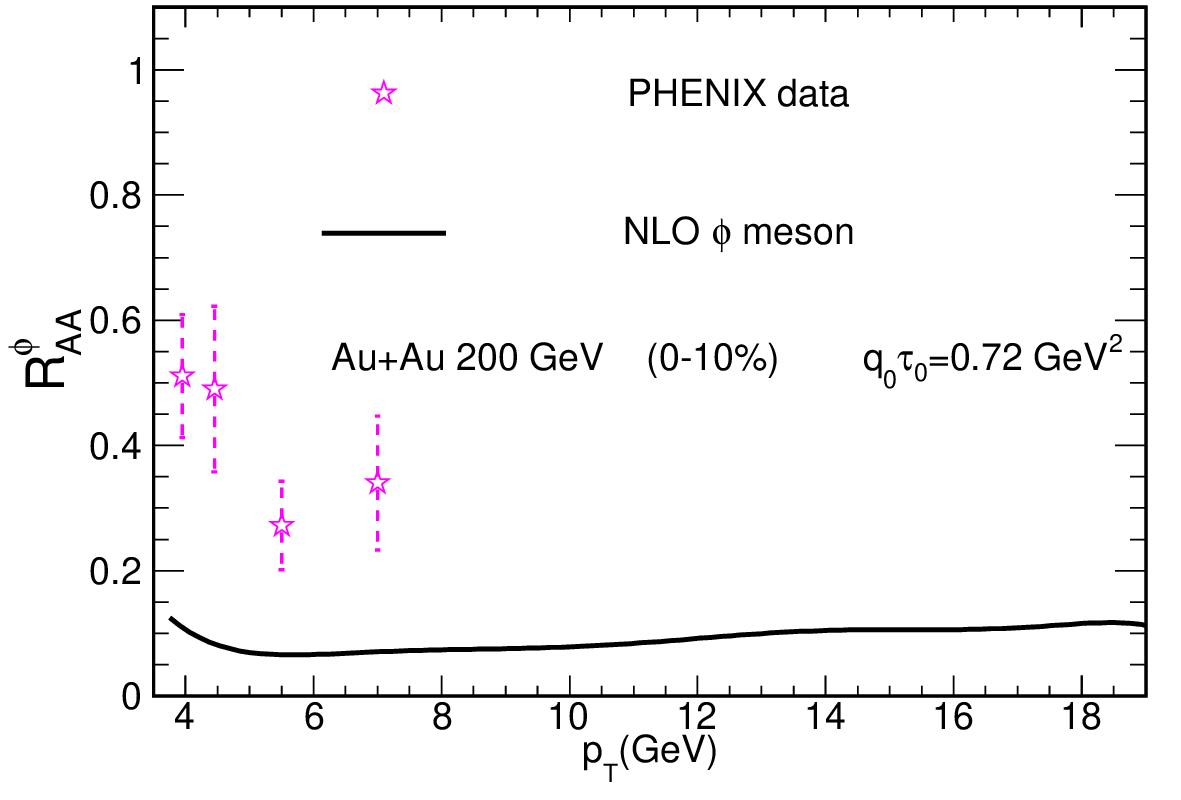}
}
\vspace*{-0.1in}
\caption{Top: The nuclear modification factor as a function of $p_T$ in Pb+Pb collisions at the LHC calculated at both LO and NLO accuracy, with ALICE data~\cite{Richer:2015vqa,Adam:2017zbf}. Bottom: The nuclear modification factor as a function of $p_T$ in Au+Au collisions at the RHIC calculated at both LO and NLO accuracy, with PHENIX data~\cite{Adare:2010pt}.
}
\label{fig:phiRHIC}
\end{center}
\end{figure}
\hspace*{-0.5in}

\hspace{0.7in}
\begin{figure}[!h]
\begin{center}
\resizebox{0.5\textwidth}{!}{%
\includegraphics{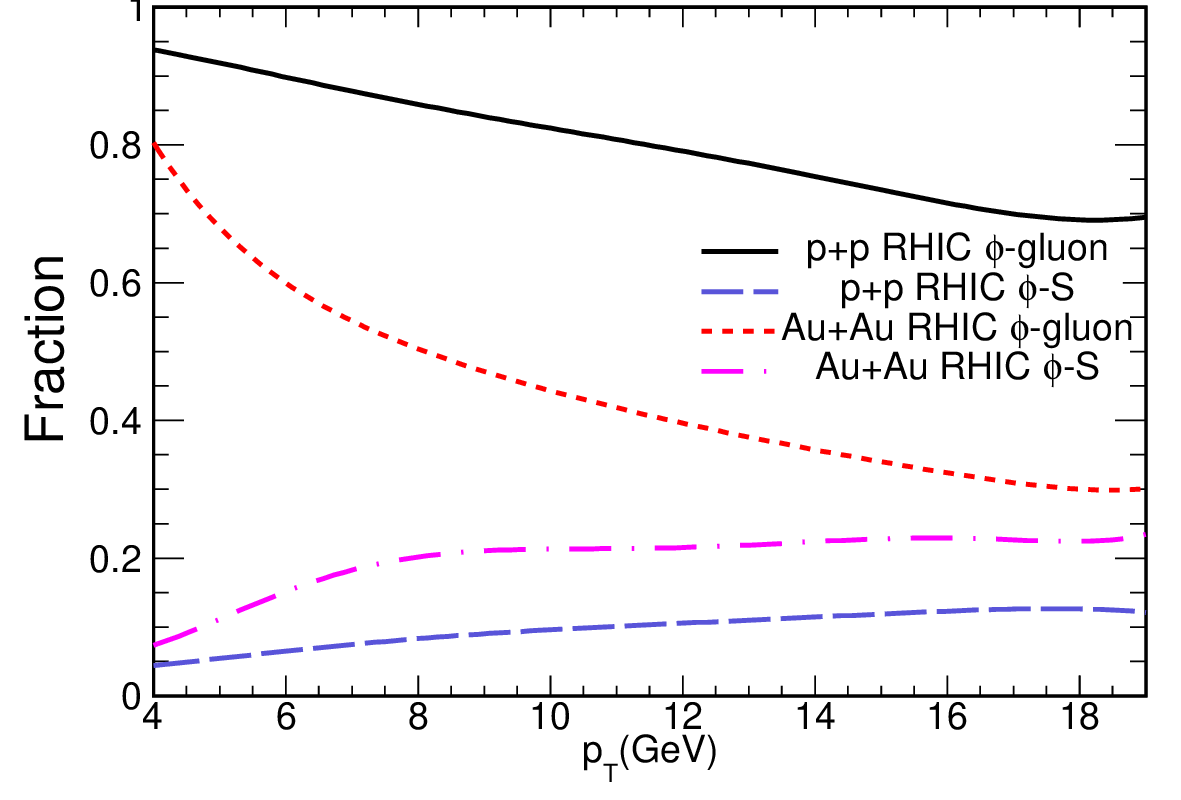}
}
\resizebox{0.5\textwidth}{!}{%
\includegraphics{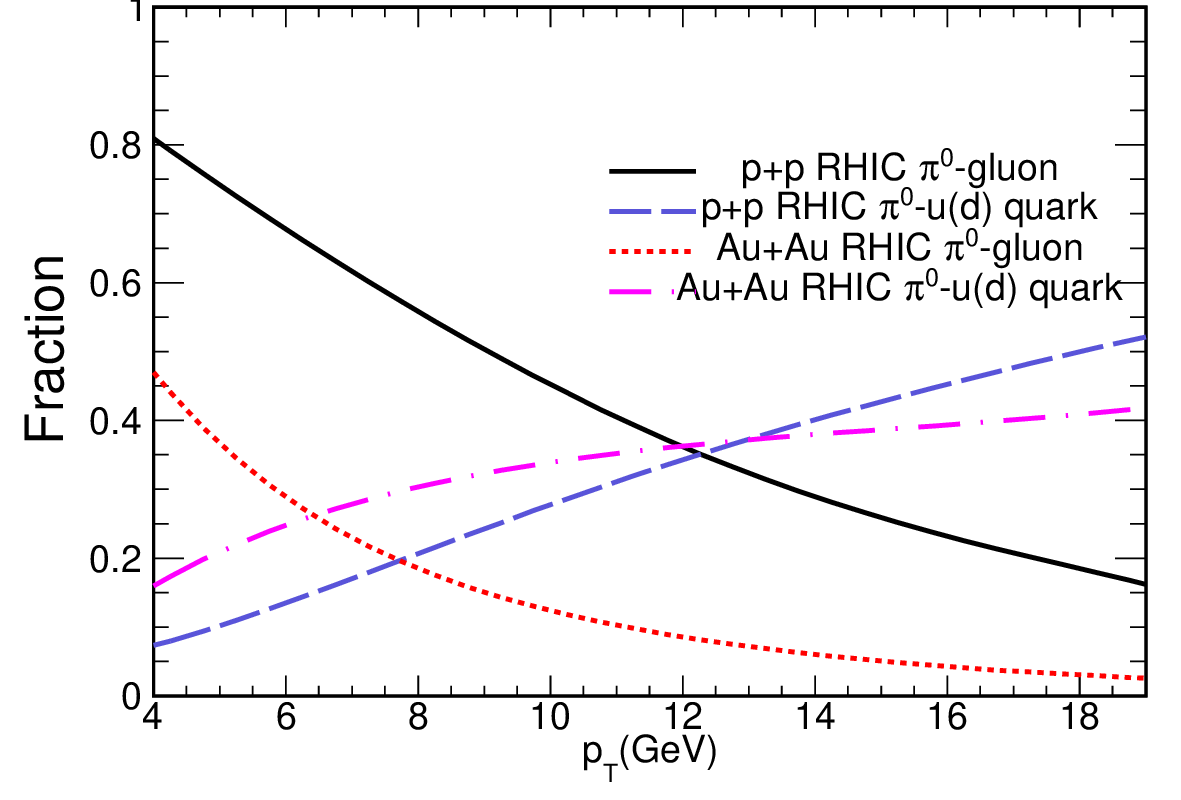}
}
\vspace*{-0.0in}
\caption{Gluon and quark contribution fractions of the total $\phi$ yield (top panel) as well as $\pi^0$ yield (bottom panel) in p+p collisions and central Au+Au collisions at the RHIC.
}
\label{fig:phifracrhic}
\end{center}
\end{figure}


A full three-dimensional (3+1D) ideal hydrodynamics description~\cite{Hirano2001,HT2002} is employed to give the space-time evolutionary information of the QCD medium such as parton density, temperature, fraction of the hadronic phase and the four flow velocity at every time step. There remains only one parameter $\hat{q}_0\tau_0$: the product of initial value of jet transport parameter $\hat{q}_0$ at the most central position in the overlap region and the initial time $\tau_0$ when the QCD medium is formed. It characterizes the overall strength of jet-medium interaction that rely on the collision energy and system, also the amount of the energy loss of the energetic jets.
To finally derive the production in A+A collisions, we replace the vacuum fragmentation functions in Eq.~(\ref{eq:ptspec}) by the initial production position and jet propagation direction averaged medium modified fragmentation functions, scaled by the number of binary nucleon-nucleon collisions at the average value of the impact parameter $b$ in $\rm A+A$ collisions. To demonstrate the medium modification of the single hadron production,  the nuclear modification factor $R_{\rm AA}$ as a function of $p_T$ is introduced to divide cross sections in $\rm A+A$ collisions by the ones in p+p, scaled by the number of binary nucleon-nucleon collisions with a chosen impact parameter $b$ as follows:
\begin{eqnarray}
R_{\rm AB}(b)=\frac{d\sigma_{AB}^h/dyd^2p_T}{N_{bin}^{AB}(b)d\sigma_{pp}^h/dyd^2p_T} \ .
\label{eq:eloss}
\end{eqnarray}
where $N_{bin}^{AB}(b)=\int d^{2}r t_{A}(r)t_{B}(|\vec b-\vec r|)$ is calculated using the Glauber model. The fixed value of impact-parameters in the calculation of the spectra and the modification factor are also determined through the Glauber geometric fractional cross sections at given centrality of the heavy-ion collisions.

We calculate the inclusive $\phi$ meson productions in nuclear nuclear collisions up to NLO at the RHIC and the LHC using this unified framework as studying $\pi^0$, $\eta$ and $\rho^0$~\cite{Chen:2010te,Chen:2011vt,Dai:2015dxa,Dai:2017tuy}. We apply the same choice of the parameter values $\hat{q}_0\tau_0$ with the initial formation time $\tau_0=0.6$~fm of the quark-gluon plasma, which has been found to give very nice descriptions of those neutral mesons in HIC. Initial-state cold nuclear matter effects is also taken into account by employing the EPS09s parametrization set of nuclear PDFs  $f_{a/A}(x_a,\mu^2)$~\cite{Eskola:2009uj}.

\section{Results and discussions}
\label{results}

In the numerical calculations, the extraction of quark jet transport coefficient $\hat{q}_0$ at the central of the most central A+A collisions at a given initial time $\tau_0$ is performed by best fitting to the PHENIX data on $\pi^0$ production spectra in $0-5\%$ Au+Au collisions at $\sqrt{s}=200$~$  \rm GeV$ which gives $\hat{q}_0=1.20\pm0.30$~$\rm GeV^2/fm$ and also fitting to the ALICE and CMS data combined on charged hadron spectra in $0-5\%$ Pb+Pb  collisions at $\sqrt{s}=2.76$~$  \rm TeV$ which gives $\hat{q}_0=2.2\pm0.5 ~\rm GeV^2/fm$ at $\tau_0=0.6 ~\rm fm/c$~\cite{Chen:2011vt,Dai:2015dxa}.  As already mentioned in Ref.~\cite{Burke:2013yra}, it is consistent with the assumption that the jet transport coefficient is proportional to the initial parton density or the transverse density of charged hadron multiplicity in midrapidity.  The charged hadron pseudorapidity density at midrapidity $dN_{ch}/d\eta  \approx1584$ in the most central $0-5\%$ Pb+Pb collisions at $\sqrt{s}=2.76$~$  \rm TeV$ is $2.3\pm 0.24$ larger than $dN_{ch}/d\eta \approx687$ in $0-5\%$ Au+Au collisions at $\sqrt{s}=200$~$  \rm GeV$. Also the ratio of the transverse hadron density in central Pb+Pb collisions at the LHC to that in Au+Au at RHIC is about $2.2\pm 0.23$ which is also very close to the value of the ratio of $\hat q_{0}^{\rm LHC}/\hat q_{0}^{\rm RHIC}\approx 1.83$. 

We firstly confront our calculation with the existing experimental data by ALICE Collaboration~\cite{Richer:2015vqa,Adam:2017zbf} in the top panel of Fig.~\ref{fig:phiRHIC}, and show $R_{\rm AA}$ as a function of $p_T$ in Pb+Pb collisions at LHC calculated in NLO by choosing $q_0\tau_0=1.32\ \rm GeV^2$ with $\tau_0=0.6$~fm. The NLO results of the $R_{AA}$ agree very well with ALICE data, which varies between $4-20$~GeV. In the bottom panel of Fig.~\ref{fig:phiRHIC} we present numerical predictions of  $R_{\rm AA}$ in Au+Au collisions at the RHIC both at NLO with the jet transport parameter $q_0\tau_0=0.72\ \rm GeV^2$ ($\tau_0=0.6$~fm), where PHENIX data~\cite{Adare:2010pt} available for a rather limited $p_T$ region ($4-7$~GeV) are also shown. Our theoretical prediction in Fig. 4 (bottom) undershoot the experimental data. In this manuscript,  the pQCD based calculation is more applicable at larger $p_T$ region, other non perturbative mechanism such as recombination in the $p_T=2-8$~GeV region is also not included.

\begin{figure}[!h]
\begin{center}
\resizebox{0.5\textwidth}{!}{%
\includegraphics{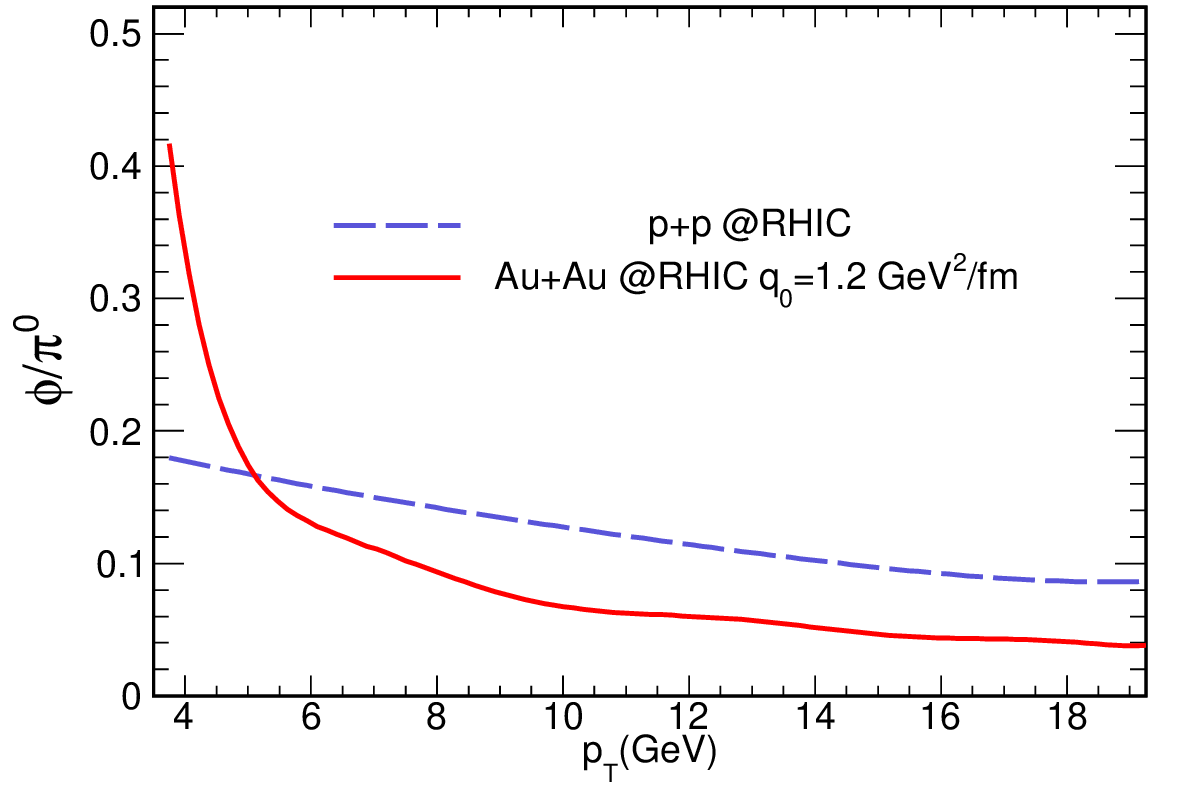}
}
\resizebox{0.5\textwidth}{!}{%
\includegraphics{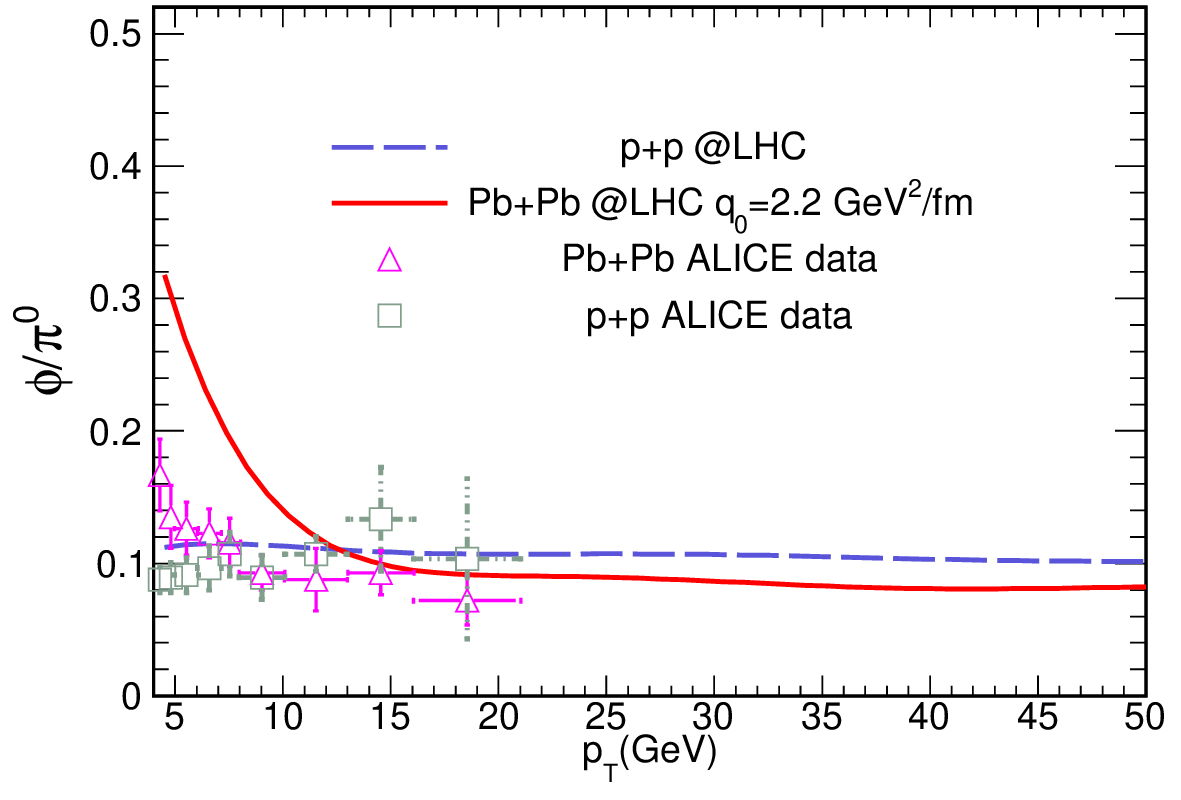}
}
\vspace*{-0.0in}
\caption{Top: $\phi/\pi^{0}$ ratio as a function of $p_T$  in p+p and Au+Au collisions at the RHIC. Bottom: $\phi/\pi^{0}$ ratio as a function of $p_T$  in p+p and Pb+Pb collisions at the LHC, comparing with ALICE data~\cite{Adam:2017zbf}. 
}
\label{fig:ratiophiAANLO}
\end{center}
\end{figure}


\begin{figure}[!t]
\begin{center}
\resizebox{0.5\textwidth}{!}{%
\includegraphics{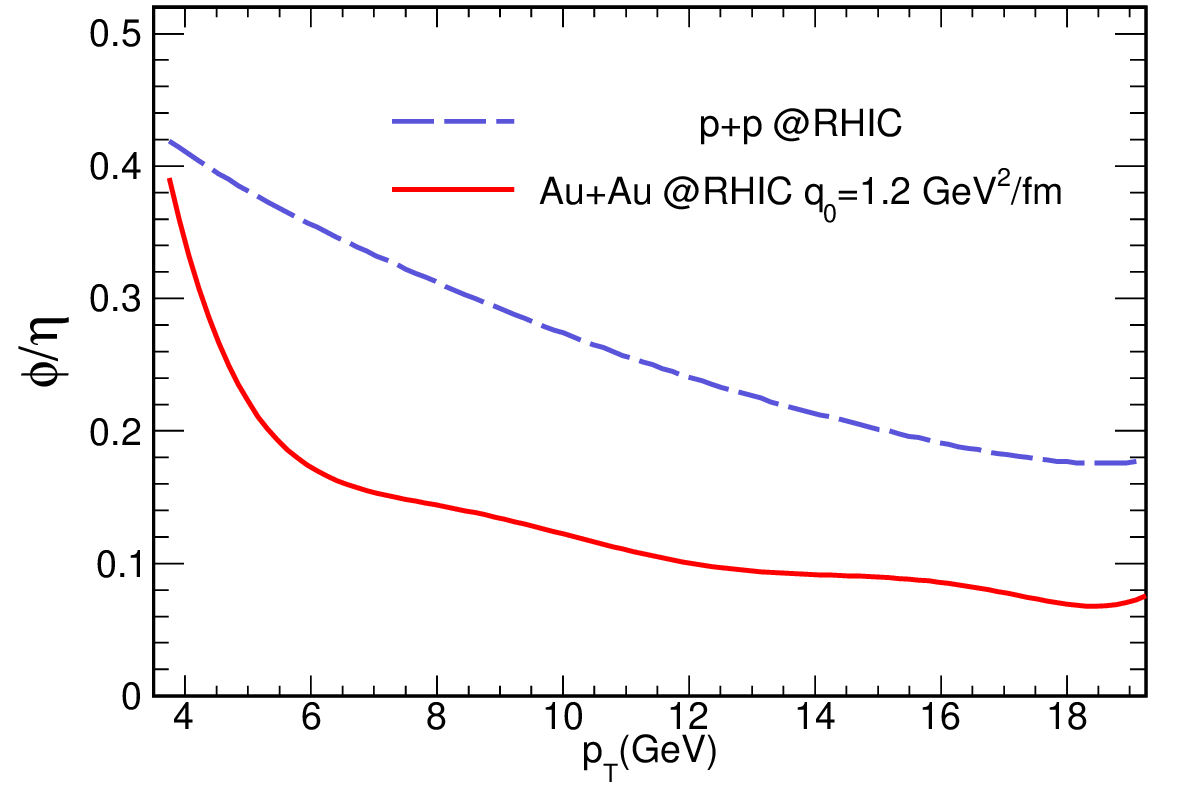}
}
\resizebox{0.5\textwidth}{!}{%
\includegraphics{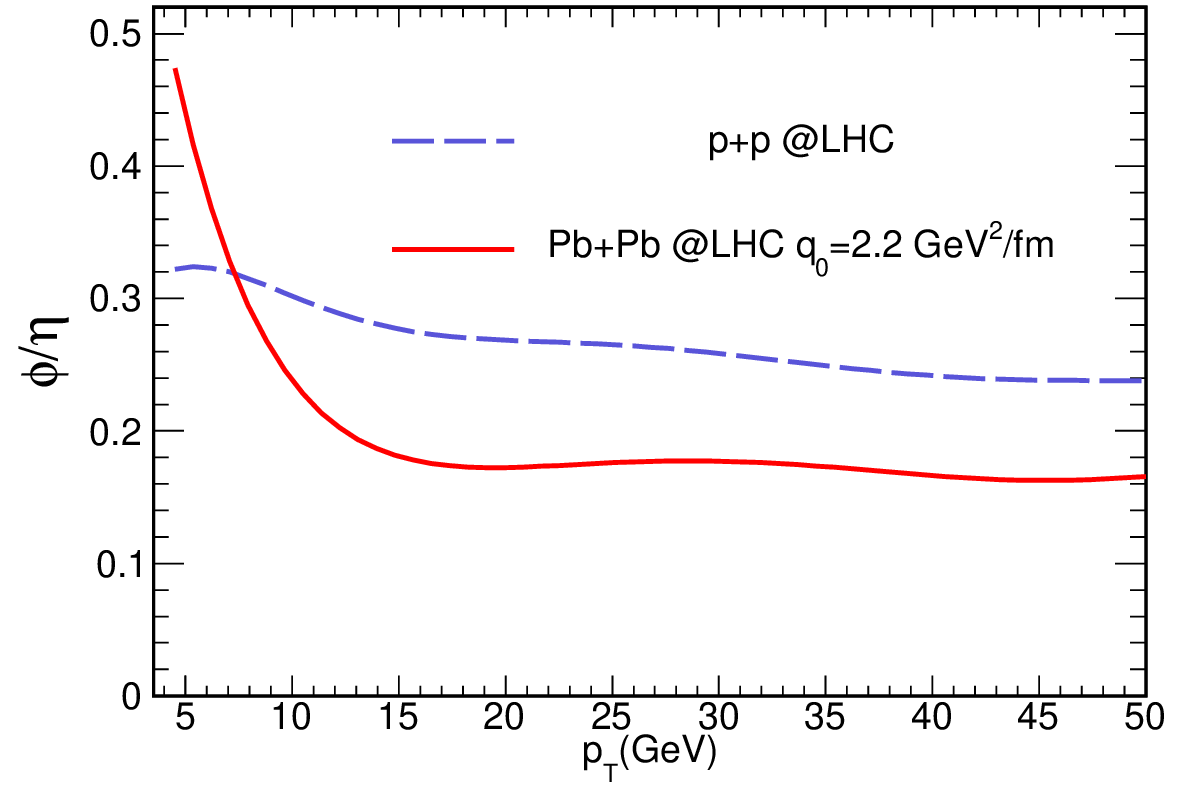}
}
\vspace*{-0.0in}
\caption{Top: yield ratio $\phi/\eta$ as a function of $p_T$  in p+p and Au+Au collisions at the at the RHIC. Bottom: $\phi/\eta$ as a function of $p_T$  in p+p and Pb+Pb collisions at the LHC.
}
\label{fig:ratiophirhiceta}
\end{center}
\end{figure}

\begin{figure}[!t]
\begin{center}
\resizebox{0.5\textwidth}{!}{%
\includegraphics{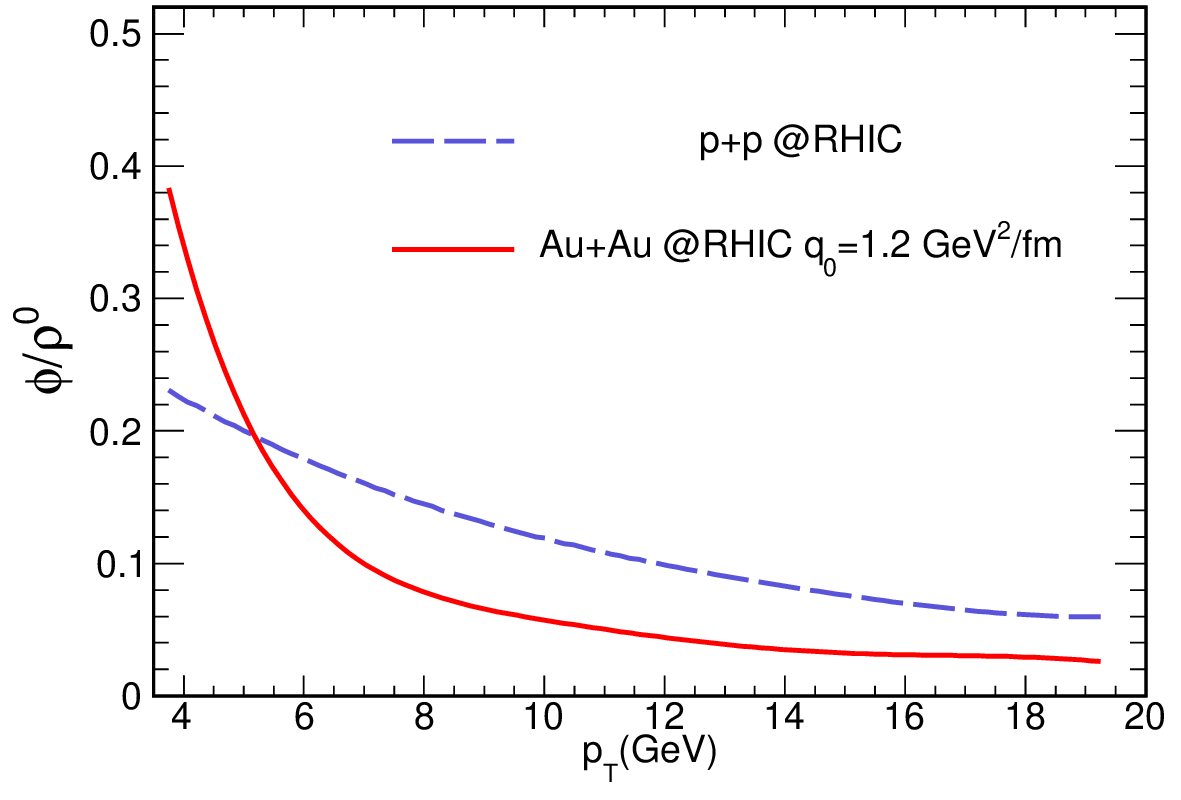}
}
\resizebox{0.5\textwidth}{!}{%
\includegraphics{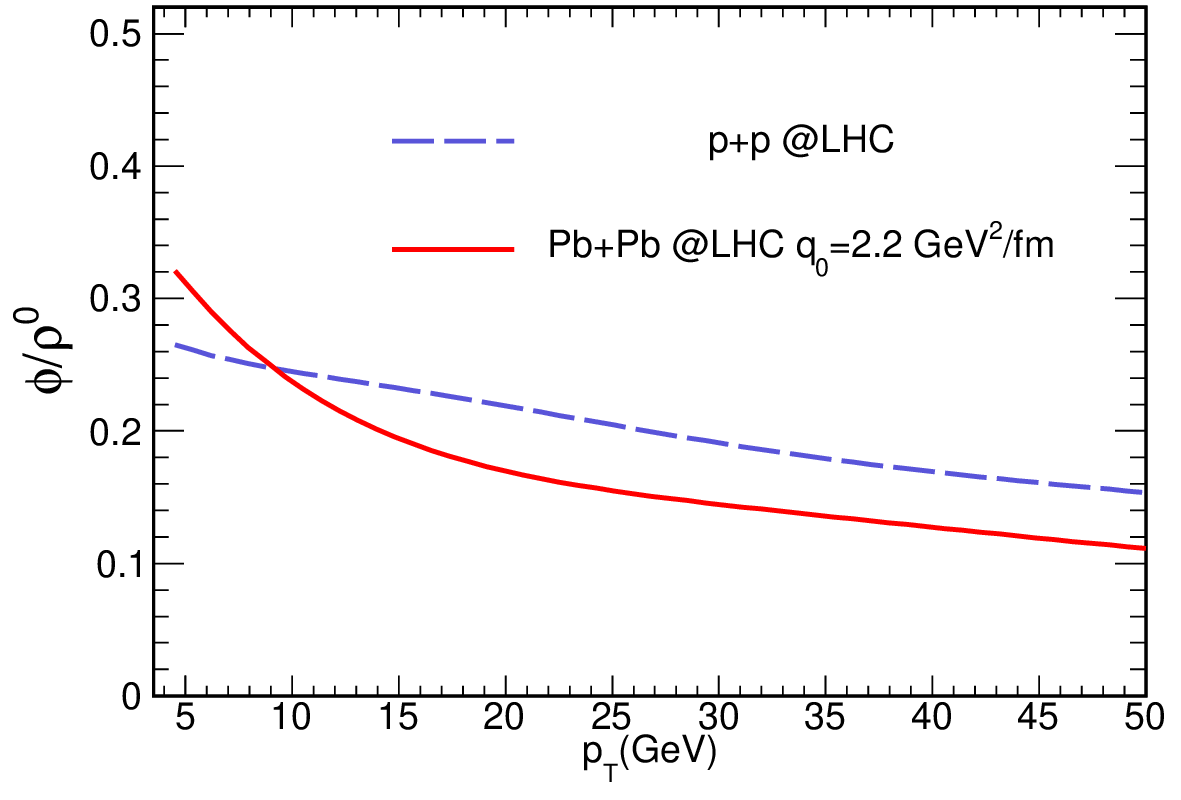}
}
\vspace*{-0.0in}
\caption{Top: $\phi/\rho^0$ as a function of $p_T$ in p+p and Au+Au collisions at the RHIC. Bottom: $\phi/\rho^{0}$ as a function of $p_T$  in p+p and Pb+Pb collisions at the LHC.
}
\label{fig:ratiophirhicrho}
\end{center}
\end{figure}

We note that $\phi$ meson production has a unique feature as compared to productions of other neutral mesons ($\pi^0$, $\eta$ and $\rho^0$). In the top panel of Fig.~\ref{fig:phifracrhic} we plot the only gluon (strange quark) fragmentating contribution fraction of $\phi$ yield in p+p collision at the RHIC. We find though in quark model $\phi$ meson is in $s{\bar s}$ state, the strange quark fragmentation only gives a $5-10\%$ contribution of the total $\phi$ meson yield, and the dominant contribution to the total $\phi$ meson yield comes from gluon fragmentation in the wide range of $p_T$ (even at the region $p_T$ $\sim20$~GeV). This feature is in striking contrast with the productions of $\pi^0$, $\eta$ and $\rho^0$. As a comparison
in the bottom panel of Fig.~\ref{fig:phifracrhic} we show 
the only gluon (strange quark) fragmenting contribution fraction of $\pi^0$ production in p+p collision at the RHIC. One can observe that 
the gluon contribution fraction to $\pi^0$ goes down under $50\%$ when $p_T\sim 9$~GeV. At high $p_T$ region, the $\pi^0$ production is dominated by light (up and down) quark contribution, which holds true also for $\eta$ and $\rho^0$ production in p+p reactions.

In A+A collisions, the parton energy loss mechanism will change the parton-jet chemistry components because a fast gluon will lose more energy in the QGP than a fast quark due to its large color-charge ($\Delta E_g/\Delta E_q =C_A/C_F=9/4$). Therefore gluon contribution fraction will be reduced in A+A collisions relative to that in p+p. In Fig.~\ref{fig:phifracrhic} we also provide the parton contribution fraction to $\phi$ meson (top panel) and to $\pi^0$ meson (bottom panel) in central Au+Au at the RHIC.
The decreasing of the gluon contribution fraction and the increasing of the quark contribution fraction observed in both cases reflect the larger energy loss suffered by the gluon-jet. For $\phi$ meson production in A+A, gluon fragmentation still gives 
$\sim 40\%$ contribution of the total yield in the intermediate $p_T$ region, and $>30\%$ at very high transverse momentum. For high transverse momentum $\pi^0$ meson production, however, because gluon contribution fraction even in p+p is not dominant, its value in A+A collisions is further suppressed and leads to a few percent at $\sim 20$~GeV. Similar trend could also be observed in $\eta$ and $\rho^0$ productions in A+A reactions.

Combining the above discussions on neutral meson productions in p+p and A+A collisions, we may reach interesting conclusions. Because at very high $p_T$ the productions of three neutral mesons ($\pi^0$, $\eta$ and $\rho^0$) are all dominated by quark fragmentation, whether in p+p or A+A collisions, the yield ratios of these three neutral mesons, for example $\eta/\pi^0$ and $\rho^0/\pi^0$, at very high $p_T$ in A+A collisions will approach to that in p+p reactions if quark FFs for these mesons at very high $p_T$ have a flat dependence on $z_h$ and the hard scale $p_T$, as seen in the theoretical calculations of the ratio $\eta/\pi^0$ in Ref.~\cite{Dai:2015dxa} and $\rho^0/\pi^0$ in
Ref.~\cite{Dai:2017tuy}, as well as related experiment observation~\cite{Adler:2006hu}. However, for $\phi$ meson production, the story will be quite different: in p+p collisions 
high $p_T$ $\phi$ meson yield is dominated by gluon fragmentations, while in A+A reactions it should be dominated by quark fragmentation because parton energy loss effect suppress the relative contribution of hard gluons; thus the yield ratios of $\phi$ meson to other neutral mesons ($\pi^0$, $\eta$ and $\rho^0$) in A+A collisions may show different behaviour with the ones in p+p reactions even at very high $p_T$ region.

In Fig.~\ref{fig:ratiophiAANLO}, we plot the yield ratio $\phi/\pi^{0}$ as functions of $p_T$  in p+p and A+A collisions with $\sqrt{s_{NN}}=200$~GeV at the RHIC,  
and with $\sqrt{s_{NN}}=2.76$~TeV at the LHC.  In Fig.~\ref{fig:ratiophirhiceta} and Fig.~\ref{fig:ratiophirhicrho} we demonstrate the yield ratios $\phi/\eta$ and $\phi/\rho^{0}$  in p+p and A+A collisions. We could observe that these three yield ratios $\phi/\pi^{0}$, $\phi/\eta$ and $\phi/\rho^{0}$ really show distinct behaviour at very high $p_T$ in A+A collisions from the ones in p+p at both the RHIC and the LHC energies, and the distinctions are more obvious at the RHIC.

We notice that the identified leading hadron production in HIC should in general be determined by three factors: the initial hard parton-jet spectrum, the parton energy loss mechanism, and parton FFs to the hadron in vacuum. These three factors are intertwined with each other. Even though leading hadrons in HIC are produced in the same scenario that the parent parton first loses its
energy in the produced QCD medium and then fragments into
a leading hadron in the vacuum with the same probabilities
governing high $p_T$ hadron production in the elementary p+p collisions, 
the high $p_T$ yield ratios of hadrons of different species in A+A may show distinct behaviour from those in p+p due to their inherited characteristic 
parton FFs. The yield ratios of $\phi/\pi^{0}$, $\phi/\eta$ and $\phi/\rho^{0}$ discussed in this manuscript demonstrated clearly this property of high $p_T$ identified hadron productions in HIC. 

In summary, we have provided the calculation and the theoretical prediction of $\phi$ meson productions in p+p and A+A collisions both at the LHC and the RHIC in the framework of pQCD for the very first time. In the calculation, higher-twist approach to the multiple scattering in the QCD medium has been used to introduce the effectively medium modified fragmentation functions to calculate the production of $\phi$ meson in A+A collisions. Due to the discovery of the gluon domination of $\phi$ production which is unlike $\pi^0$, $\eta$ and $\rho$, we find the calculated yield ratio of $\phi/\pi^0$, $\phi/\rho^0$ and $\phi/\eta$ are not shown the coincidence between  A+A and p+p, which is displayed among the ratios of light quark dominated mesons: $\rho^0/\pi^0$, $\eta/\pi^0$. Therefore, the ratio of $\phi$ meson to the other light quark mesons such as $\pi^0$, $\eta$ and $\rho^0$,  will provide an interesting probe of the colour charge sensitivity of jet quenching.
\vspace*{.3cm}

{\bf Acknowledgments:} We thank Prof. Xin-Nian~Wang for his helpful discussion. This research is supported by the MOST in China under Project No. 2014CB845404, and by NSFC of China with Project Nos. 11435004, 11322546, and 11521064. 

\vspace*{-.6cm}

%

\end{document}